\documentclass[a4paper,twocolumn,showpacs,preprintnumbers,amsmath,amssymb,nofootinbib,floatfix]{revtex4}
\usepackage{hyperref}
\input psfig.sty
\usepackage{picinpar}
\usepackage{graphicx}

\preprint{GCON-2016-P02}

\begin{document}

\title{Non-parametric reconstruction of cosmological matter perturbations}

\author{J. E. Gonz\'alez\footnote{E-mail: javierernesto@on.br}}

\author{J. S. Alcaniz\footnote{E-mail: alcaniz@on.br}}

\author{J. C. Carvalho\footnote{E-mail: carvalho@dfte.ufrn.br}}

\affiliation{Departamento de Astronomia, Observat\'orio Nacional, 20921-400, Rio de Janeiro - RJ, Brasil}

\date{\today}

\begin{abstract}

Perturbative quantities, such as the growth rate ($f$) and index ($\gamma$), are powerful tools to distinguish  different dark energy models or modified gravity theories even if they produce the same cosmic expansion history. In this work, without any assumption about the dynamics of the Universe, we apply a non-parametric method to current measurements of the expansion rate $H(z)$ from cosmic chronometers and high-$z$ quasar data and reconstruct the growth factor and rate of linearised density perturbations in the non-relativistic matter component. Assuming realistic values for the matter density parameter $\Omega_{m0}$, as provided by current CMB experiments, we also reconstruct the evolution of the growth index $\gamma$ with redshift. We show that the reconstruction of current $H(z)$ data constrains the growth index to $\gamma=0.56 \pm 0.12$ (2$\sigma$)  at $z = 0.09$, which is in full agreement with the prediction of the $\Lambda$CDM model and some of its extensions.

\end{abstract}


\maketitle

\section{Introduction}

The cosmic acceleration, first inferred from type Ia supernovae (SNe Ia) observations in the late 1990s~\cite{Riess, Perlmutter}, cannot be explained in the framework of the General Relativity (GR) with the material content of the Universe satisfying the strong energy condition.
This in turn poses a major challenge for theoretical physics and has led  physicists to hypothesize the existence of dark energy (DE), a negative pressure energy component which dominates the energy content of the universe at present (for a review, see~\cite{review}). 

In order to achieve cosmic acceleration, GR equations for a homogeneous and isotropic universe require $w < -(\Omega_{m0}/3\Omega_{\rm{DE,0}} + 1/3)$, where $w$ is the ratio between the dark energy pressure to its energy density, and $\Omega_{m0}$ and $\Omega_{\rm{DE,0}}$ stand for the present-day density parameters of the clustered matter and of the dark energy, respectively. Since very little is known about the nature of this DE field (e.g., it is unclear if its energy density is in fact time-independent), alternative explanations for cosmic acceleration have been suggested. These are mainly based on modification of gravity at large scales and examples of them include scalar-tensor gravity~\cite{Scalar}, $f(R)$ theories~\cite{fR,Starobinsky2007},  higher dimensional braneworld models~\cite{DGP}, among others (an extensive list of DE models and modified gravity theories is discussed in~\cite{Copeland_DDE} and references therein).

From the observational point of view, it is well known that the accuracy of present background observations -- e.g., measurements of luminosity and angular diameter distances and of the cosmic expansion rate -- is not enough to distinguish between DE models and scenarios of modified gravity. In reality, even for highly accurate data, it is not possible to decide which one gives the best description of the Universe because different models can produce the same cosmic expansion and, therefore, the same background observables. However, the existing degeneracy at the background level can be lifted by the study of the growth of matter density perturbation, $\delta$~\cite{Wang2008}. As well known, in theories of modified gravity the growth rate for $\delta$ is usually different from that predicted by general relativistic models, with the effective gravitational constant $G_{eff}$, which appears in the source term driving the evolution of $\delta$,  changing significantly relative to the Newton's gravitational 
constant $G$, usual 
in the GR regime.  

On the other hand, given the large number of competing cosmological models and the inherent difficulties of distinguishing between them, parametric and non-parametric methods have been developed with the aim of obtaining independent information about the physics behind cosmic acceleration from observations (see \cite{CPL-Polarski,CPL-Linder,BA,Sendra,Vitenti,Montiel,Rasmussen,Seikel,Seikel2012,Shafieloo2006,Shafieloo2012a,Shafieloo2012b,Li} and references therein). In this paper, we apply a non-parametric method, namely, Gaussian processes (GP)~\cite{Rasmussen,Seikel}, to a set of observational data to reconstruct 
the {{growth factor, $g$, and rate}}, $f$, of linear perturbations and the {{growth index}}, $\gamma$, following closely the treatment developed in Ref.~\cite{Sahni2009}.  In our analysis we use cosmological model-independent measurements of the cosmic expansion rate $H(z)$ lying in the redshift range $0.070 \leq z \leq 2.34$. Currently, most of the $H(z)$ data available come from measurements of age differences of the so-called cosmic chronometers, i.e., passively evolving galaxies at different $z$~\cite{Jimenez-Loeb}, whose uncertainties are around 10\%-15\%~\cite{Jimenez03,Simon05,Stern10,Moresco12,Moresco:2015cya}. Estimates of the expansion rate have also been obtained from the three-dimensional correlation function of the transmitted flux fraction in the Ly$\alpha$-forest of high-$z$ quasars, as reported in Refs.~\cite{Busca,Debulac,Font-Ribera}. In particular, the application of this latter technique to a large sample of quasars provided measurements of $H(z)$ within $\sim 3\%$ accuracy at $z \simeq 2.3$, which imposes 
tight 
bounds on cosmological parameters when combined with current $H_0$ measurements and other cosmological data sets (see \cite{Farooq} for a recent analysis). From a subsample of 
the currently available $H(z)$ data, we reconstruct perturbative quantities from background observations and investigate possible tensions between current data and the DE models predictions.

This paper is organised as follows: in Sec \ref{MPE} we  summarise the treatment developed in Ref. \cite{Sahni2009} introducing the basic expressions that govern the matter perturbation growth and the related quantities.  We discuss the observational data  and the non-parametric method used  to reconstruct the cosmic history in Sec. \ref{DaHPR}. In Sec. \ref{R} we present the reconstructed functions of the perturbative quantities and discuss their compatibility with the standard cosmological description. We end this paper by summarizing the main conclusions in Sec. \ref{C}.

\section{Matter Perturbation Equations}
\label{MPE}

The scalar perturbations of a flat Friedmann-Lemaitre-Robertson-Walker (FLRW) metric are characterised, in the longitudinal gauge, by the line element
\begin{equation}
 ds^2=(1+2\Phi)dt^2-(1-2 \Psi)a^2(t)d \vec{x} ^2\;,
\end{equation}
where $\Phi$ and $\Psi$ are the gauge invariant potential and curvature perturbation, respectively. In GR these quantities are equal if we neglect any anisotropic  stress which could, for instance, be produced by primordial neutrinos\footnote{If we consider anisotropy stress, then $\nabla^2 (\Phi + \Psi) \neq 0$, mimicking some models of modified gravity. In our analysis we assume that the dark energy component does not have anisotropic stress and does not couple to matter.}. On sub-Hubble scales, the potencial satisfies the Poisson equation
\begin{equation}
 \nabla^2 \Phi =4\pi G a^2 \rho_m \delta\;,
\end{equation}
where
\begin{equation}
 \delta(\vec{x},t)\equiv \frac{\rho(\vec{x},t)-\rho(t)}{\rho(t)}
\end{equation}
is the non-relativistic matter density contrast.

Assuming GR and a background filled with matter and an unclustered component of DE covariantly conserved, the linearised matter density contrast 
satisfies the second order differential equation 
\begin{equation}
\label{2ode}
\ddot{\delta}+2H\dot{\delta}-4 \pi G \rho_m \delta = 0\;.
\end{equation}

The covariant conservation of the matter energy-momentum tensor implies $\rho_m \propto (1+z)^{3}$. Using this result and the definition of the dimensionless physical distance given by
\begin{equation}
D=H_0 \int_t^{t_0}\frac{dt}{a(t)}=H_0\int_0^z \frac{dz_1}{H(z_1)}\;,
\end{equation}
Eq. (\ref{2ode}) can be rewritten as \cite{Sahni2009}:
\begin{equation}
\label{22ode}
\left(\frac{\delta '}{1+z(D)} \right)'=\frac{3}{2}\Omega_{m0}\delta\;,
\end{equation}
where a prime denotes derivative with respect to $D$ and $\Omega_{m0}$ is the matter density parameter at the present time. 
The solution of the Eq (\ref{22ode}) can be written in terms of a set of integral equations as follows \cite{Sahni2009,SahniStarobinsky}:
\begin{subequations}
\begin{eqnarray}
\delta(D)&=&1+\delta_0 '\int_0 ^D[1+z(D_1)]dD_1  \\ &&
 +\frac{3}{2}\Omega_{m0}\int_0^D[1+z(D_1)]\left(\int_0^{D_1}\delta(D_2)dD_2  \right)dD_1 \nonumber \;,
 \label{22odes}
\end{eqnarray}
\begin{eqnarray}
\delta'(D)&=&\delta_0 '[1+z(D)] \\ && +\frac{3}{2}\Omega_{m0}[1+z(D)] 
\int_0^{D}\delta(D_1)dD_1 \nonumber \;.
\end{eqnarray} 
\end{subequations}
In order to solve the previous set of integral equations one needs to assume a value for $\Omega_{m0}$. In our analysis, we adopt two different estimates of this quantity, as provided by Planck and WMAP collaborations (see Sec. \ref{R}). Also, we fix the two integration constants to obtain a unique solution of the Eq. (\ref{22ode}).  The first one $\delta_0=\delta(z=0)$ is implicitly fixed by  Eq. (\ref{22odes}) since the solution is normalised to its value today, i.e., $\delta_0=1$ whereas the second one, $\delta'_0=\delta'(z=0)$, is fixed with the requirement that the behaviour of the density contrast at high redshift must be proporcional to $a$.  In practice, however, it is easier to fix $\delta'_0$  analysing the growth factor, defined as 
\begin{equation}
g(z)\equiv(1+z)\delta(z)\;.
\end{equation}

A unique solution for Eq. (\ref{22ode})  implies that the cosmic expansion history in GR determines univocally the matter density contrast as pointed out in Refs. {\cite{Wang2008,Nesseris2015}}.  
Therefore, we can perform an indirect determination of $\delta$ reconstructing the Hubble parameter from the $H(z)$ data. 
As shown in Ref. \cite{Starobinsky1998} the inverse problem, i.e., the determination of cosmic expansion as a function of the density contrast, $H(z)=H(\delta(z))$, has an analytical solution given by
\begin{equation}
 H^2(z)=3\Omega_{m0}H^2_0 \frac{(1+z)^2}{(d\delta/dz)^2} \int_z^\infty \frac{\delta \lvert d\delta/dz \rvert}{1+z} dz.
\end{equation}
The growth rate of linear perturbations is defined as 
\begin{equation}
 f(z)\equiv\frac{d\ln \delta}{d\ln a}=-\frac{(1+z)H_0}{H(z)}\frac{\delta'}{\delta}\;.
 \label{fdef}
\end{equation}
Note that the values of the growth rate  obtained by solving the above equation via reconstruction of $H(z)$ constitute an independent estimate {{of this quantity as inferred from the matter power spectrum or weak gravitational lensing data \cite{Reyes2010}}}. A tension between them would be an evidence of non-standard cosmology where the Eq. (\ref{2ode}) is not valid. If this is the case, it would imply that:

\begin{itemize}
 \item The Universe is not correctly described by a flat FLRW metric. For instance, in the case of a non-flat and inhomogenous universe the Poisson equation is modified, as shown in Ref. \cite{Marteens}.
 
 \item The evolution of the matter density is not proportional to  $(1+z)^3$. It implies that the matter energy-momentum  tensor is not covariantly conserved. 
Typical examples are models with decaying of dark energy into dark matter or vice versa \cite{Rsousa}. 
 
 \item 
 The GR  is not valid and it needs to be modified or  the DE is clumping and its effect has to be taken into account. 
 In  both cases, it would be possible to define an effective gravitational function, $G\rightarrow G_{eff}$ which can depend on the scale and time \cite{GannoujiPolarski, carroll, Tsujikawa}.
\end{itemize}

The growth index is written as~\cite{Peebles,Lahav,Wang1998}
\begin{equation}
\gamma = \frac{d\ln{f(z)}}{d\ln{\Omega_{m}(z)}}\;,
 \label{fapprox}
\end{equation}
where
\begin{equation}
 \Omega_m(z) \equiv \frac{\Omega_{m0}(1+z)^3H_0^2}{H^2(z)}\;,
 \label{om}
\end{equation}
is the matter density parameter as a function of redshift.  As mentioned earlier, the growth rate and growth index constitute key quantities to distinguish between modified gravity and DE models 
\cite{Wang2008,GannoujiPolarski}. 
For instance, $\gamma=6/11$ for $\Lambda$CDM \cite{Wang1998,LinderCahn}, $\gamma=11/16$ for the so-called DGP scenarios \cite{LinderCahn} and lies in the interval $0.40<\gamma<0.43$ \cite{GannoujiPolarski} for the $f(R)$ model proposed in Ref. \cite{Starobinsky2007}.  For non clustering  DE models, $\gamma$ is related to a slowly evolving DE equation-of-state $w$ through 
 $\gamma \simeq \frac{3(w-1)}{6w-5}$
~\cite{Wang2008}.

\begin{figure*}[t]
\includegraphics[width = 8.5cm, height = 6.3cm]{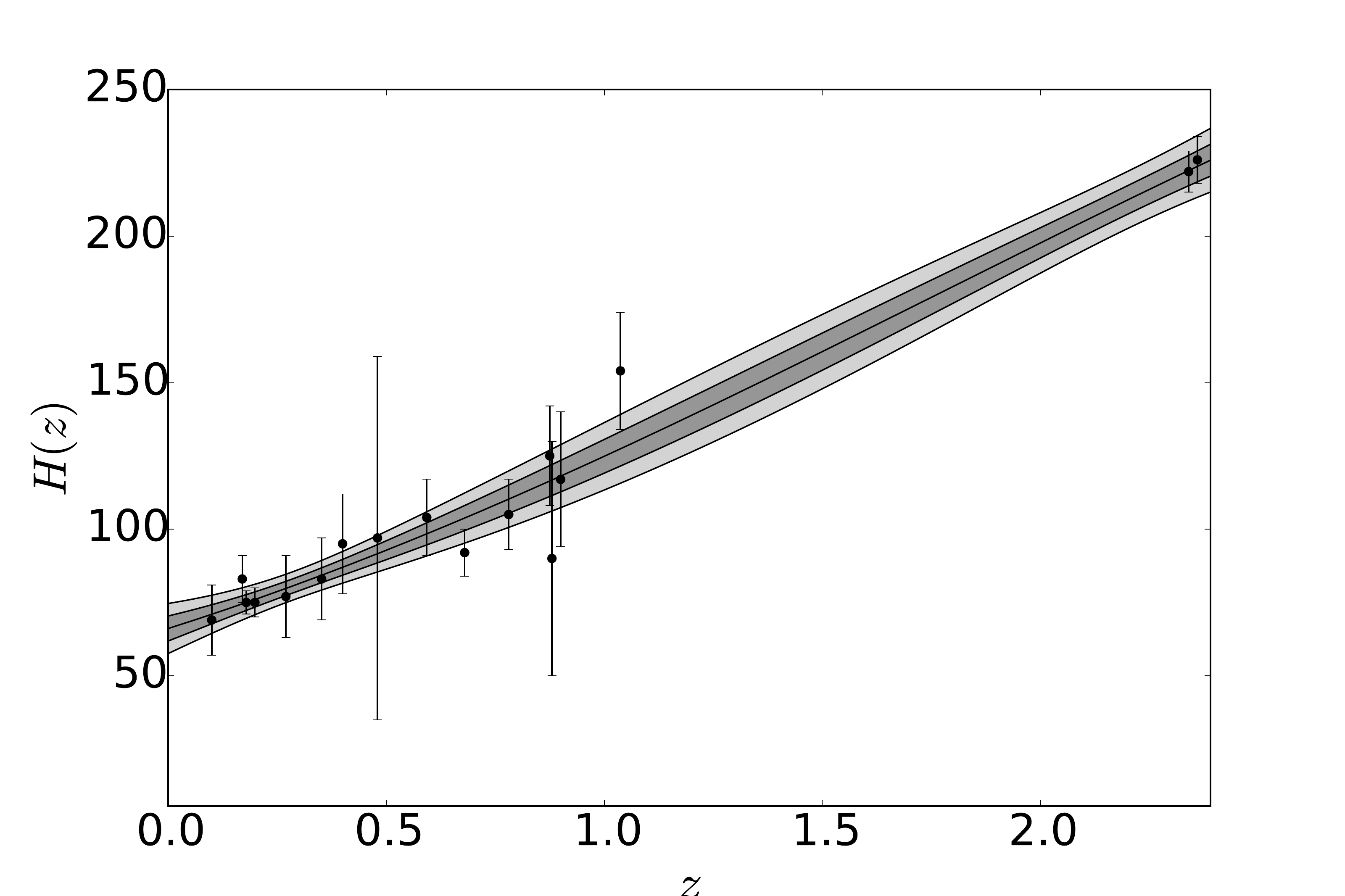}
\hspace{0.1cm}
\includegraphics[width = 8.5cm, height = 6.3cm]{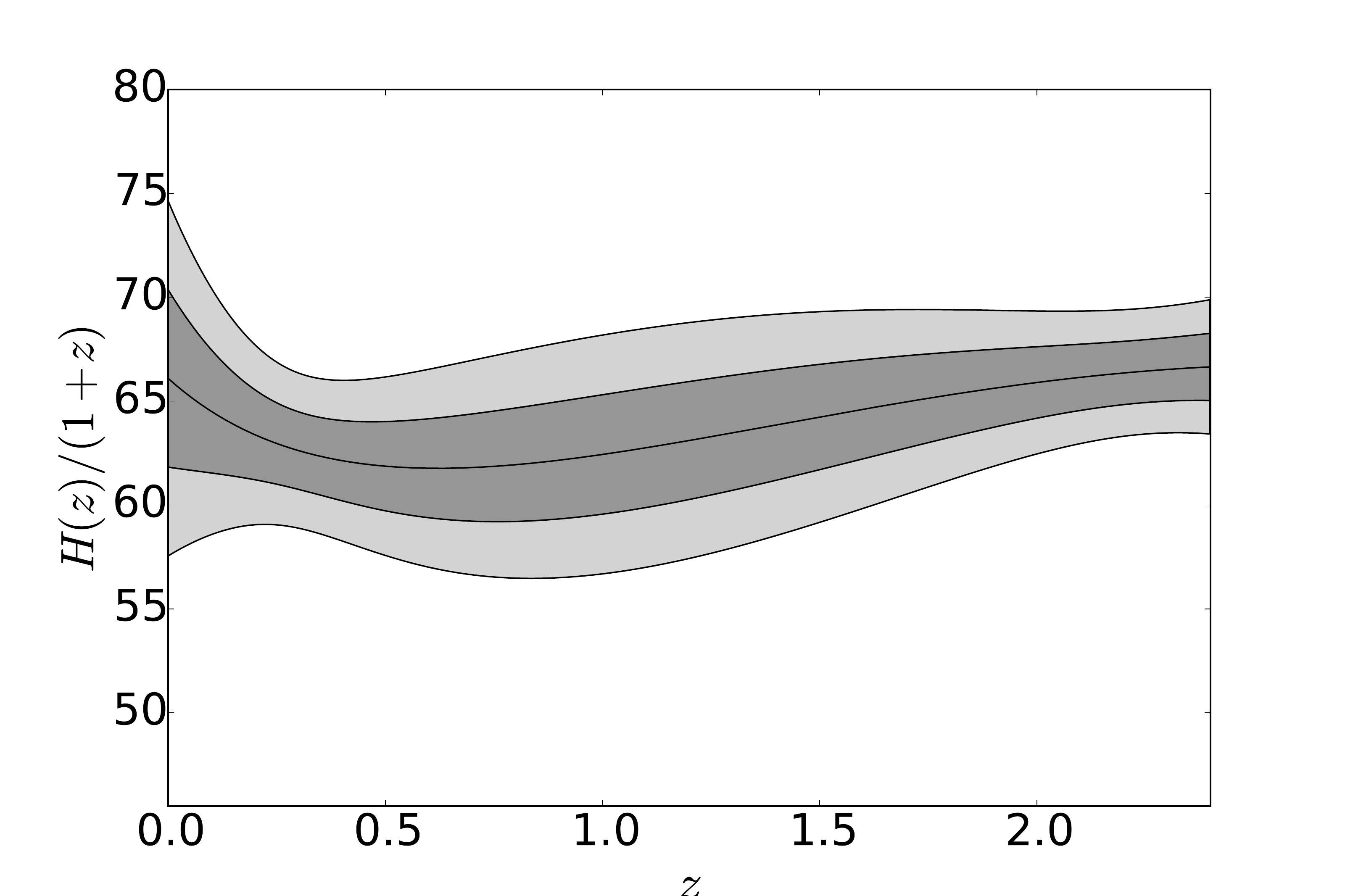}
\caption{Reconstruction of the cosmic expansion (in $\rm{{km}/{s}/Mpc}$) via Gaussian Processes from cosmic chronometer and high-$z$ quasar data. 
The black solid line corresponds to the GP reconstruction whereas the shaded regions to the 1$\sigma$ and 2$\sigma$ confidence intervals. The data points represent the observational data displayed in Table \ref{Hdados}. b) The quantity $H(z)/(1 + z)$ as a function of $z$.}
\label{Hzfig}
\end{figure*}

\begin{figure*}[t]
\includegraphics[width = 8.5cm, height = 6.3cm]{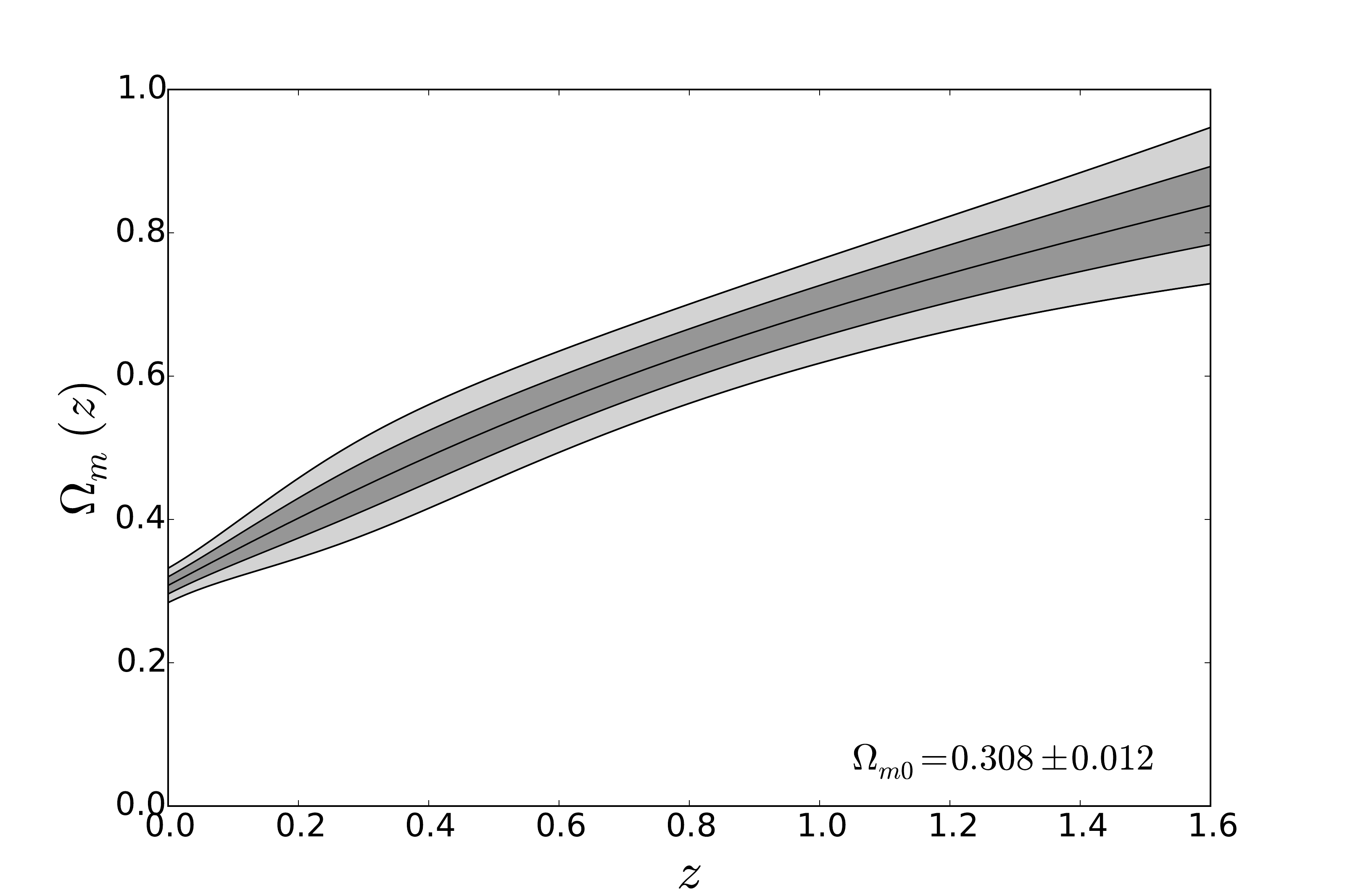}
\hspace{0.1cm}
\includegraphics[width = 8.5cm, height = 6.3cm]{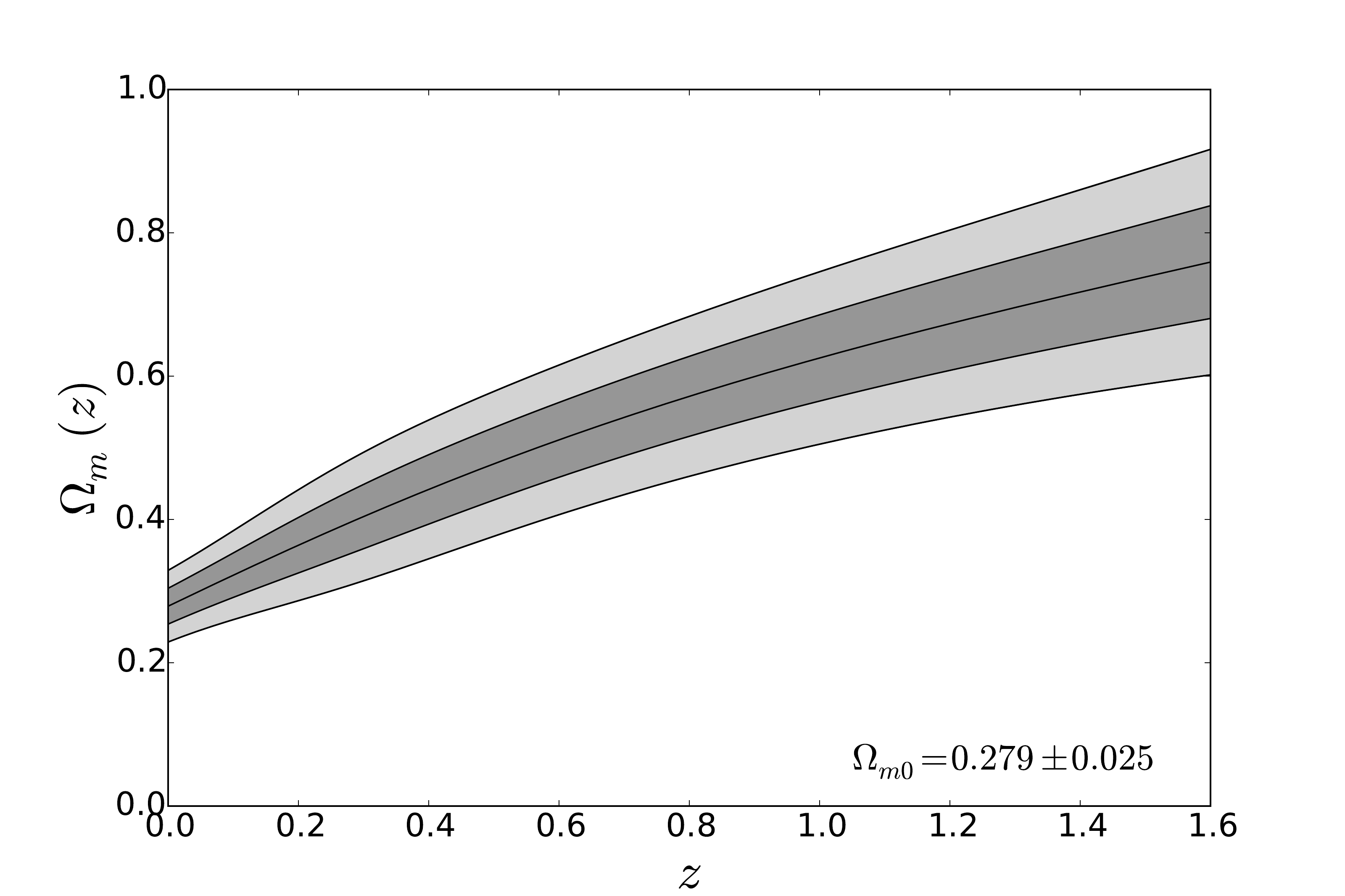}
\caption{The evolution of the matter density parameter calculated using the reconstruction of $H(z)$ shown in Fig. \ref{Hzfig}(a) and current estimates of $\Omega_{m0}$ from the Planck  (a) and the WMAP collaborations (b). The shaded regions correspond to 1$\sigma$ and 2$\sigma$ confidence intervals.}
\label{Omfig}
\end{figure*}

\section{Data and Hubble Parameter Reconstruction}
\label{DaHPR}

\subsection{Data}
\label{D}
Currently, there are different approaches to measure directly the expansion rate of the Universe.  One of them is based on the determination of the age difference between passively evolving galaxies at approximately the same redshift and known as cosmic chronometers~\cite{Jimenez-Loeb}. This information (redshift and age) provides $H(z)\simeq - \Delta z/\Delta t (1+z)$, where $\Delta t$ is the difference between the age estimates of two galaxies whose redshifts differ by $\Delta z$. Presently, there are 23 measurements of $H(z)$ using the differential age approach~\cite{Jimenez03,Simon05,Stern10,Moresco12,Moresco:2015cya}.  This method is  cosmological model-independent but there can be dependence on stellar population synthesis models at high redshift. In our analysis, we follow the arguments of Ref. \cite{Licia} and consider 15 $H(z)$ measurements up to $z\simeq1.04$. We also increase slightly (20\%) the error bar of the highest-$z$ point to account for the uncertainties of the stellar population synthesis 
models. We 
complement our sample with two high-$z$ quasar data at $z=2.34$ \cite{Debulac} and $z=2.36$ \cite{Font-Ribera} which were obtained by determining the BAO scale from the correlation function of the  Ly$\alpha$ forest systems (see \cite{Busca} for more details). The data set used in our analysis is shown in Table I.

 \begin{table}
	\begin{center}
 		\begin{tabular}{lcc}
 		\hline\hline
 $z$ & $H_{obs}(z)$ [km s$^{-1}$ Mpc$^{-1}$] & Ref. \\
 \hline
 0.100 & 69 $\pm$ 12 & \cite{Simon05} \\
 0.170 & 83 $\pm$ 8 & \cite{Simon05} \\
  0.179 & 75 $\pm$ 4 & \cite{Moresco12}\\
  0.199 & 75 $\pm$ 5 & \cite{Moresco12} \\
  0.270 & 77 $\pm$ 14 & \cite{Simon05} \\
  0.352 & 83 $\pm$ 14 & \cite{Moresco12} \\
  0.400 & 95 $\pm$ 17 & \cite{Simon05} \\
  0.480 & 97 $\pm$ 62 & \cite{Stern10} \\
  0.593 & 104 $\pm$ 13 & \cite{Moresco12} \\
  0.680 & 92 $\pm$ 8 & \cite{Moresco12} \\
  0.781 & 105 $\pm$ 12 & \cite{Moresco12}\\
  0.875 & 125 $\pm$ 17 & \cite{Moresco12} \\
  0.880 & 90 $\pm$ 40 & \cite{Stern10} \\
  0.900 & 117 $\pm$ 23 & \cite{Simon05} \\
  1.037 & 154 $\pm$ 20 & \cite{Moresco12} \\
 2.34 & 222 $\pm$ 7 & \cite{Debulac} \\
 2.36 & 226 $\pm$ 8 & \cite{Font-Ribera}\\
 \hline\hline
 		\end{tabular}
 	\end{center}
 	\caption{Measurements of the expansion rate from 15 cosmic chronometer systems and  two high-$z$ quasar data used in the analysis. }
 	\label{Hdados}
 \end{table}

\subsection{Gaussian Processes}
 A Gaussian process is the generalisation  of a Gaussian distribution of a random variable to a function space. It constitutes a powerful method to reconstruct the expected function that describes the behaviour of a given data. GP use a few assumptions about the  characteristics of the expected function $W(z)$, e.g., a correlation between the $W(z)$ and $W(z')$ values, $z$ and $z'$ being different points (see Ref. \cite{Seikel} and references therein for a complete review of the method). In any case, the reconstruction can be made  without assuming a model or a parametric function to describe the data.  

This method has shown great success, being applied to  reconstruct  several cosmological quantities like the DE equation of state~\cite{Seikel}, the deceleration parameter, the duality-distance parameter~\cite{Santos,Y.Zhang} and to infer the Hubble constant \cite{BustiH0, Li, Licia}. In GP the variation of the expected function in two different points is not independent and it is characterised by a covariance function $k(z,z')$.  The covariance depends on a set of hyperparameters (non model parameters) which determine the correlation between the $W(z)$ and $W(z')$ values. The fact that the $H(z)$ parameter must be infinitely differentiable allows us to choose a Gaussian covariance given by:
\begin{equation}
 k(z,z')=\sigma \exp{\left( -\frac{(z-z')^2}{2l^2}\right)},
\end{equation}
where $\sigma$ and $l$ are the so-called hyperparameters related to typical changes in the function values and to the length scale between two points $z$ and $z'$, respectively. In order to perform the non-parametric reconstruction of the cosmic expansion history we use the code Gaussian Processes in Python\footnote{http://www.acgc.uct.ac.za/$\sim$seikel/GAPP/index.html} applied to the $H(z)$ data presented in Table \ref{Hdados} (we refer the reader to \cite{Rasmussen} for more details on GP).

\begin{figure*}[t]
\includegraphics[width = 8.5cm, height = 6.3cm]{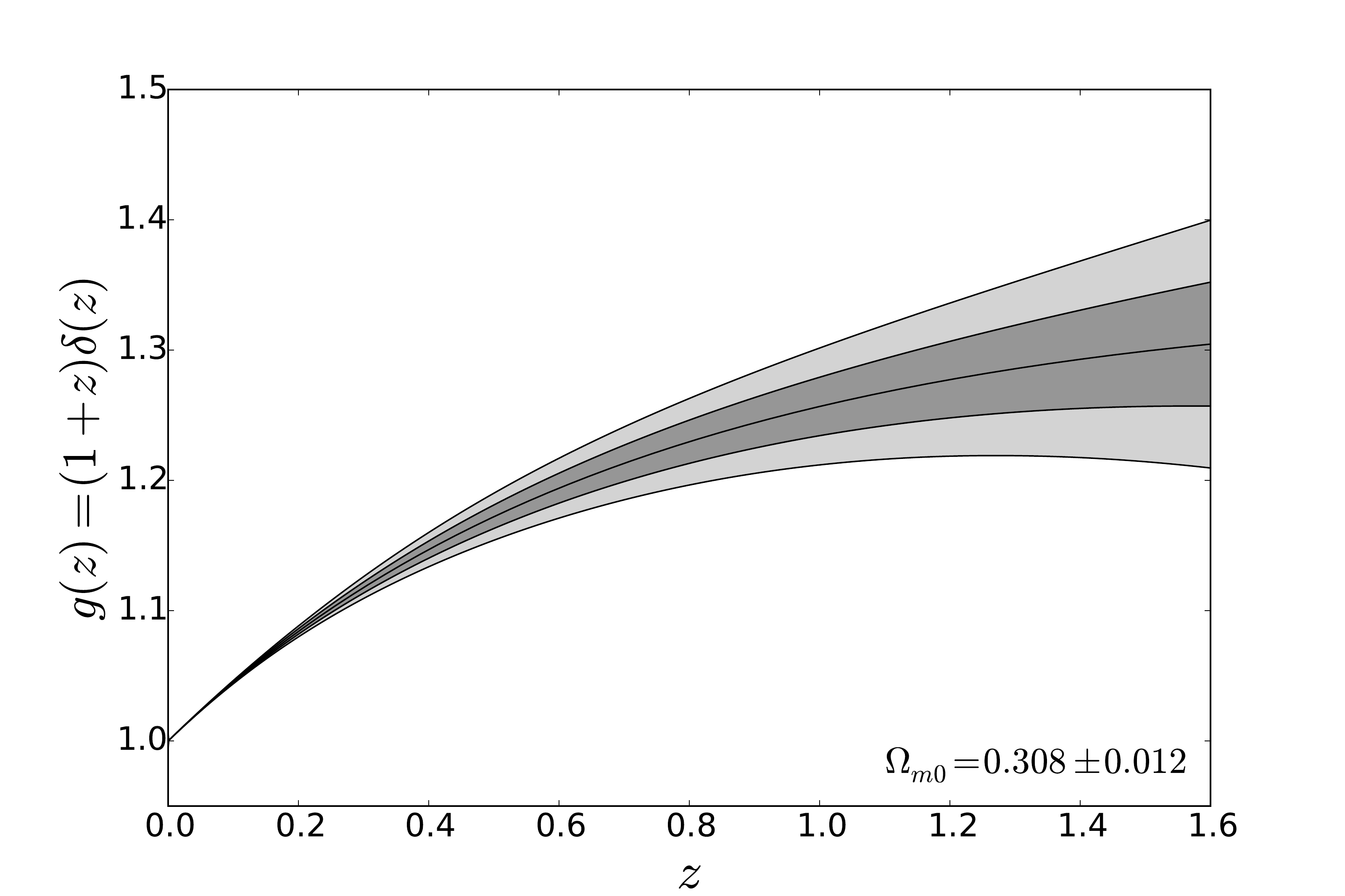}
\hspace{0.1cm}
\includegraphics[width = 8.5cm, height = 6.3cm]{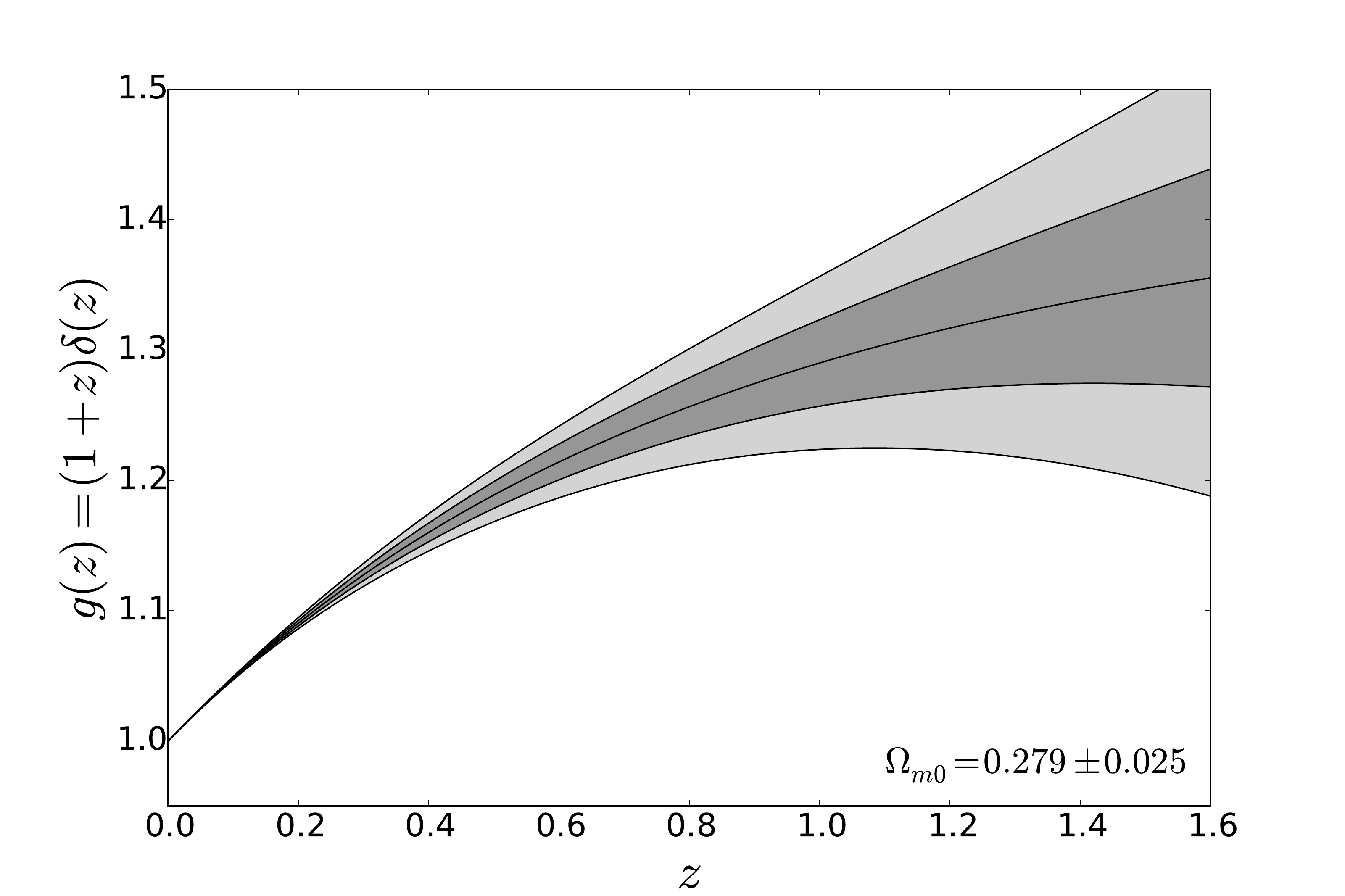}
\caption{a) The growth factor on sub-Hubble scale obtained solving Eq. \ref{22odes} using the Planck 2015 $\Omega_{m0}$ value. The solid line corresponds to the reconstruction whereas the shaded regions represent 1$\sigma$ and 2$\sigma$ confidence levels. b) The same as in the previous panel assuming the value of $\Omega_{m0}$ given by the WMAP collaboration.}
\label{gz}
\end{figure*}

\begin{figure*}[t]
\includegraphics[width = 5.7cm, height = 5.5cm]{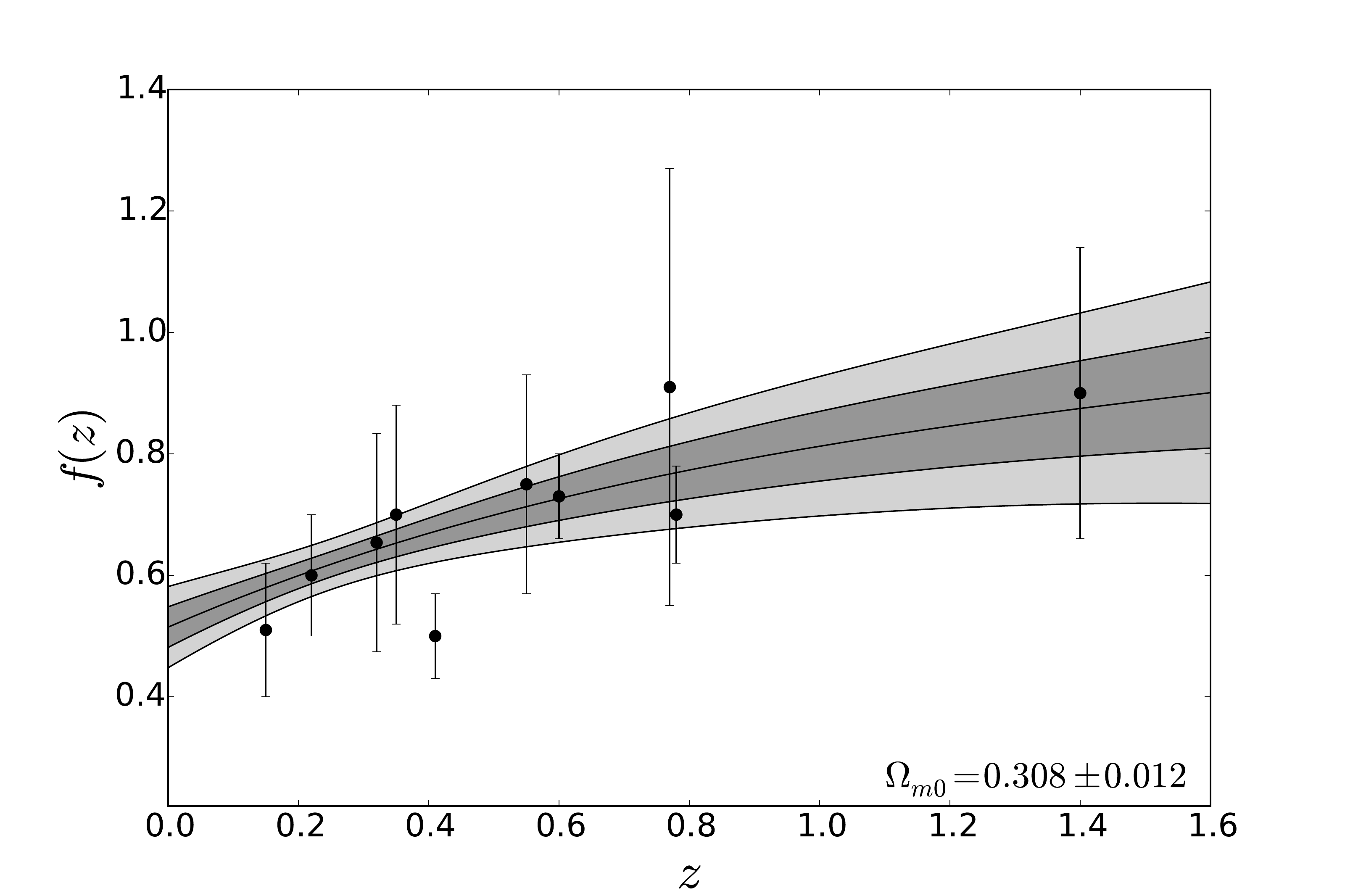}
\hspace{0.1cm}
\includegraphics[width = 5.7cm, height = 5.5cm]{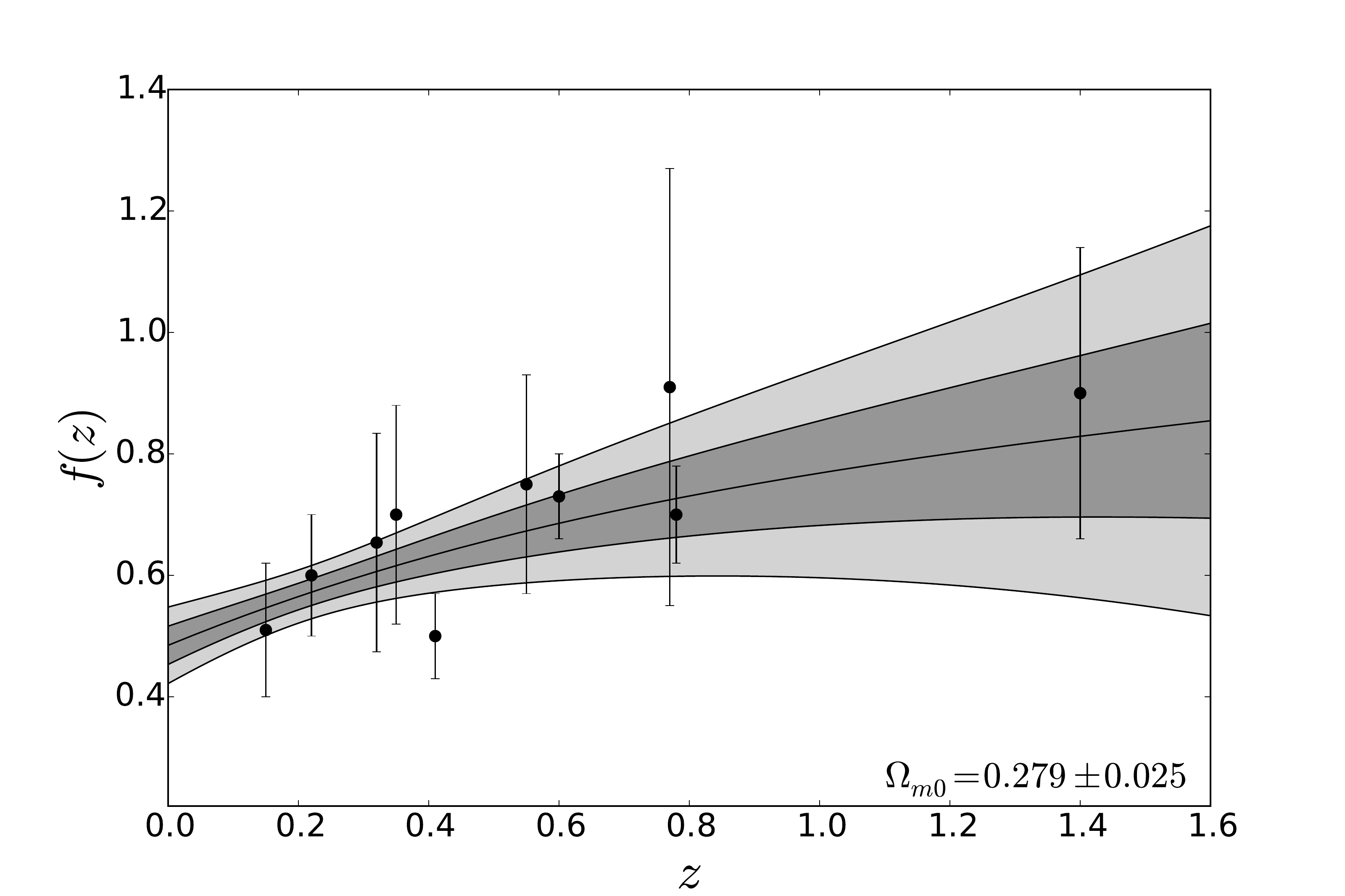}
\hspace{0.1cm}
\includegraphics[width = 5.7cm, height = 5.5cm]{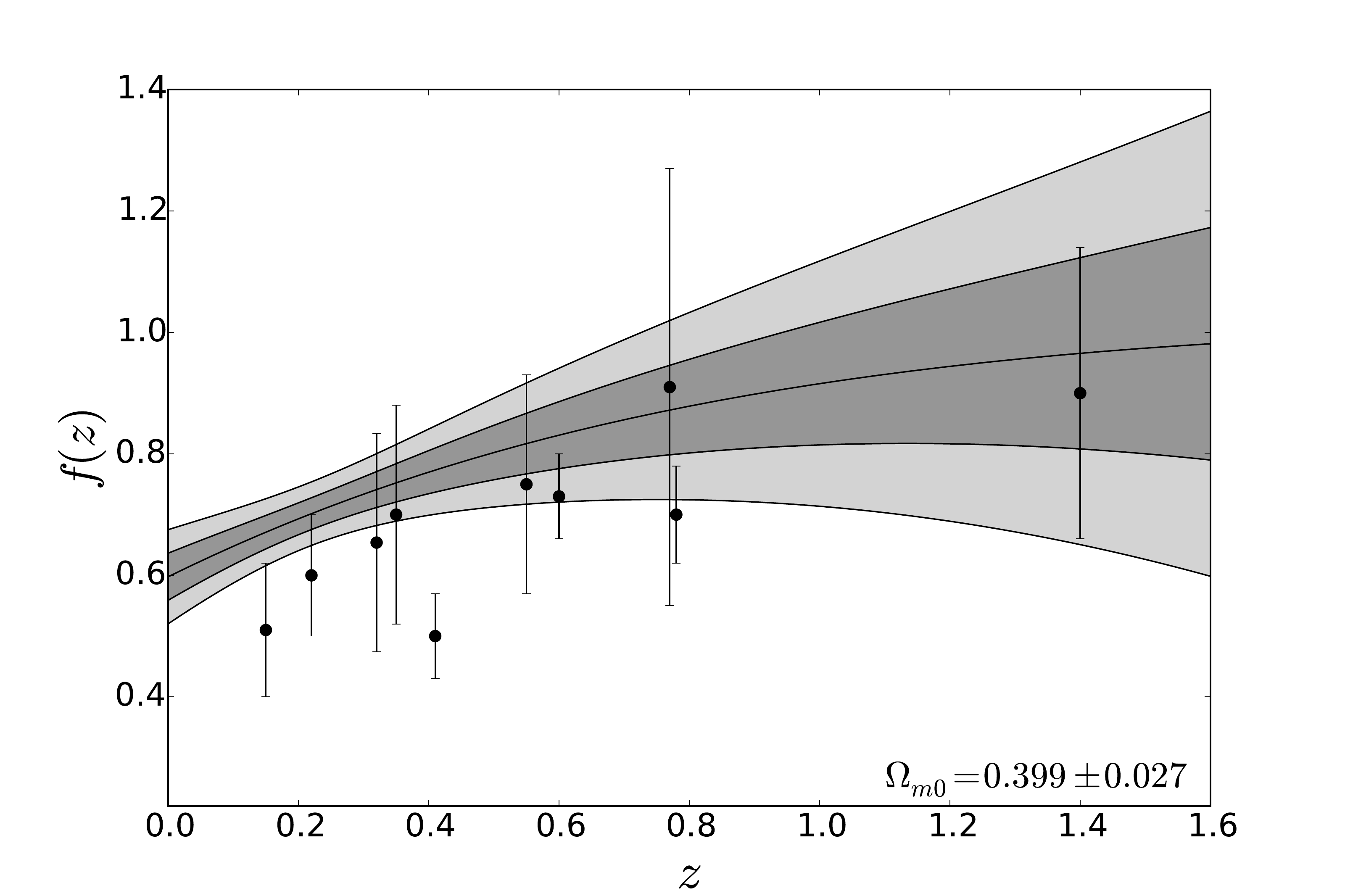}
\caption{The growth rate of the matter perturbation. The solid line corresponds to the reconstruction  whereas the shaded regions  represent 1$\sigma$ and 2$\sigma$ confidence levels. a) The growth rate  obtained assuming the value of $\Omega_{m0}$  given by the Planck  collaboration. b) The same as in the previous panel assuming the WMAP  $\Omega_{m0}$ value. c) The same as in the previous panels assuming $\Omega_{m0} = 0.399 \pm 0.027$, as obtained from SNe Ia observations~\cite{trotta}. The  data points were taken from Table II of Ref. \cite{GuptaTabela}}
\label{fz}
\end{figure*}

 \begin{figure*}[t]
\includegraphics[width = 8.5cm, height = 6.3cm]{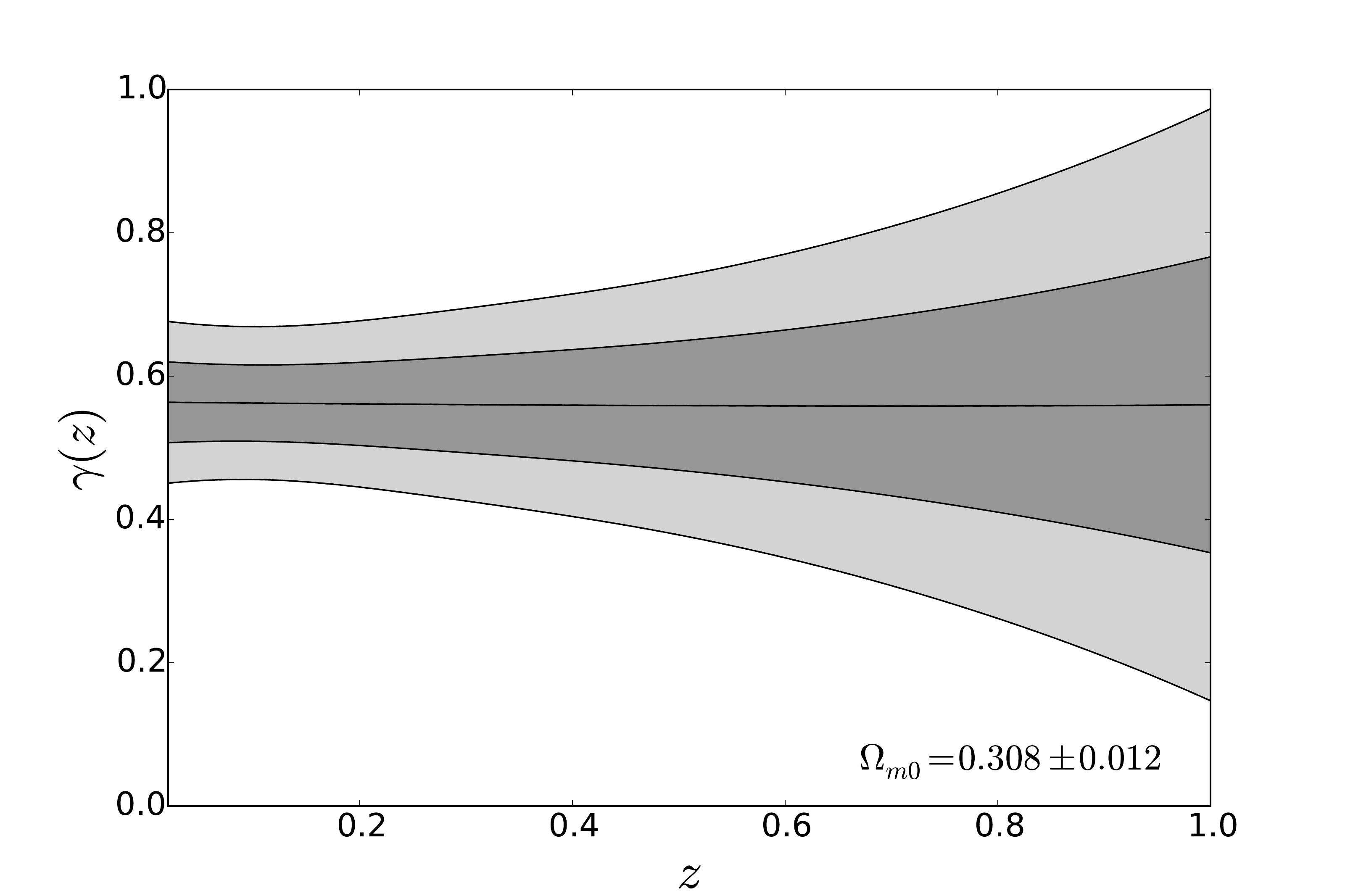}
\hspace{0.1cm}
\includegraphics[width = 8.5cm, height = 6.3cm]{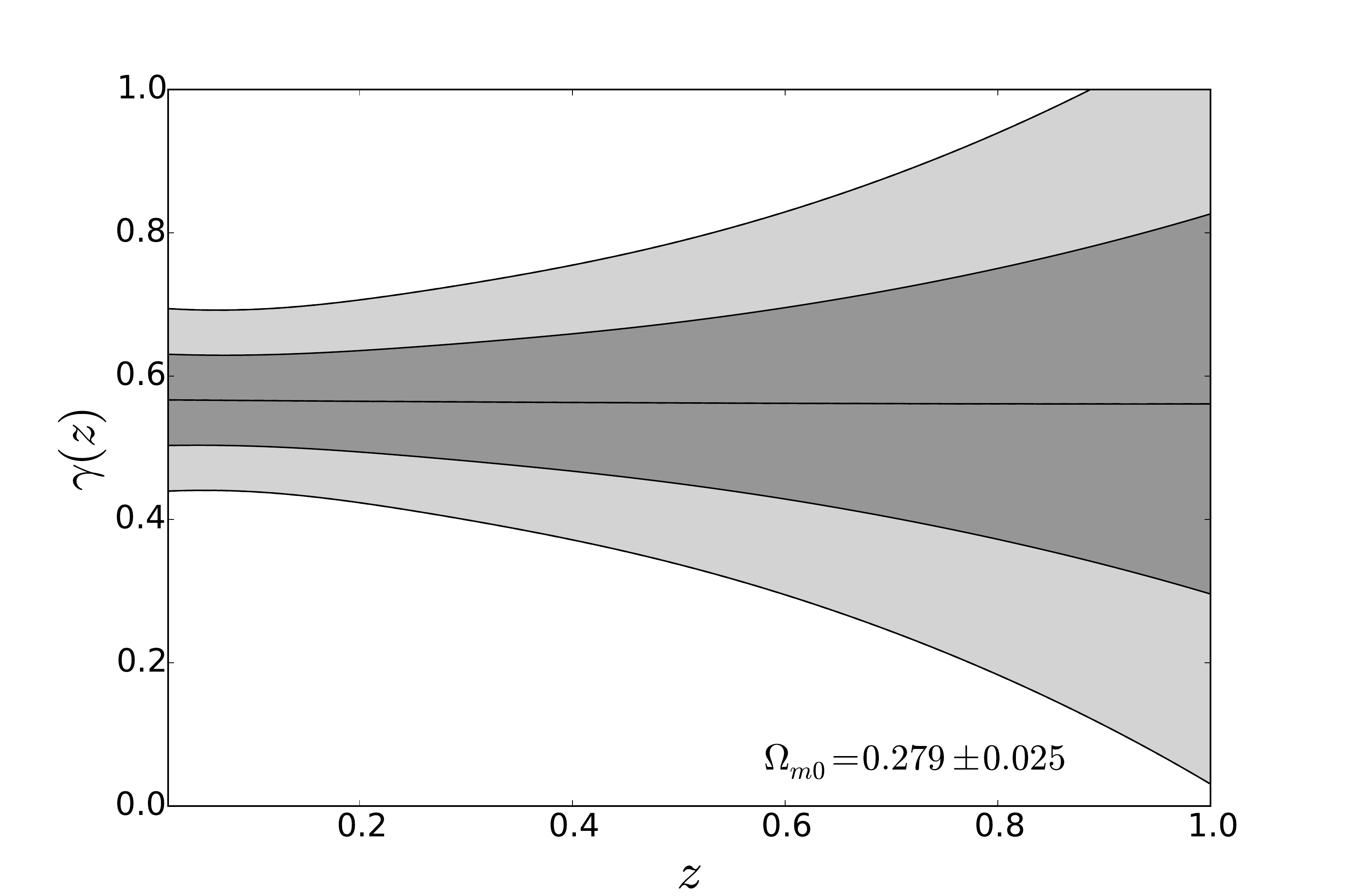}
\caption{The growth index $\gamma(z)$ of matter perturbation. The solid line corresponds to the reconstruction from GP whereas the shaded regions  represent 1$\sigma$ and 2$\sigma$ confidence levels. a) The growth index obtained assuming the Planck  $\Omega_{m0}$ value. b) The same as in the previous panel assuming the value of $\Omega_{m0}$ given by the WMAP collaboration.}
\label{gammaz}
\end{figure*}
 
\section{Results}
\label{R}

For the best-fit values of the GP hyperparameters, the reconstructed $H(z)$ function is shown in Fig. \ref{Hzfig}(a) along with the data points used in our analysis (Table I). We also plot the quantity $H(z)/(1 + z)$ as a function of $z$ in Fig. \ref{Hzfig}(b), which shows a minimum at $z \simeq 0.62$, corresponding to the recent deceleration/acceleration transition. The evolution of $\Omega_m(z)$ [Eq. (\ref{om})] is shown in Fig. \ref{Omfig}(a) adopting $\Omega_{m0}=0.308 \pm 0.012$, as given by the Planck collaboration~\cite{Planck} and in Fig. \ref{Omfig}(b) adopting $\Omega_{m0}=0.279 \pm 0.025$, as given by the WMAP collaboration~\cite{WMAP9}. 

After reconstructing the expansion rate $H(z)$ we calculate the density contrast solution (\ref{22odes}) in an iterative way. We perform the calculation of $\delta(z)$ considering the two different values of the present-day matter density parameter as mentioned above. The unique free parameter in Eq. (\ref{22odes}) is $\delta'_0$ and we need to find an appropriated value for it. From the perturbation theory we expect $\delta \propto a = 1/(1+z)$ at high-$z$. Therefore, the growth factor $g(z)=(1+z)\delta(z)$ must satisfy the condition $g(z)\rightarrow$ const. in this limit.  We fix the value of $\delta'_0$ when we reach the required behaviour of $g(z)$ close to $z = 2.34$, the highest redshift of our data set.  For the Planck and WMAP values of $\Omega_{m0}$, respectively, we estimate $\delta'_0=0.515 \pm 0.003$ and $\delta'_0=0.485 \pm 0.003$ at 1$\sigma$. The calculated $g(z)$ functions are shown in  Fig. \ref{gz}. We note that they are very similar to the ones obtained in Ref.~\cite{Sahni2009} using a non-
parametric smooth reconstruction from SNe Ia data.
 As expected (see Sec. \ref{MPE}), the $g(z)$  reconstructed function depends significantly on the present-day value of the matter density parameter assumed in the analysis.

In the reconstruction of the growth rate $f(z)$, $\delta'_0$ plays an important role because these quantities are related through $f(0)=-\delta'_0$. The resulting reconstruction, the growth rate as a function of $z$, is shown in the Fig. \ref{fz}. For comparison, we also display current measurements of this quantity, as discussed in Ref.~\cite{GuptaTabela}. The clear compatibility between the reconstruction of $f(z)$ from cosmic chronometers data and the  measurements of the growth rate from galaxy surveys can be seen as a measure of consistency of the theoretical treatment introduced in Ref.~\cite{Sahni2009} as well as of the non-parametric method of reconstruction used in the present analysis.  More importantly, for the values of $\Omega_{m0}$ given by the current CMB experiments, the results of Fig. \ref{fz} show a good agreement with the standard cosmological description, i.e., a general relativistic universe described by the FLRW line element and whose matter content is covariantly conserved (see Sec. 
II). 
Note, however, that this conclusion may change if one considers values of the matter density parameters far from the current CMB interval. This is clearly seen in Fig. \ref{fz}c which assumes $\Omega_{m0} = 0.399 \pm 0.025$, as obtained from a recent analysis of type Ia supernova data (assuming the $\Lambda$CDM model)~\cite{trotta}. Quantitatively speaking, a fit of the $f(z)$ data  to the $f(z)$ reconstructed curves provides $\chi^2 = 7.51$ and $\chi^2 = 5.20$ for the values of $\Omega_{m0}$ displayed in Panels 4a and 4b, respectively, and $\chi^2 = 25.80$ for the SNe Ia value considered in Panel 4c.

Finally, we also calculate the growth index $\gamma$ using the reconstructed function of $f(z)$ and the CMB values of $\Omega_{m0}$ discussed above. At $z = 0$, we found $\gamma_0=0.56 \pm 0.12$ ($2\sigma$) and  $\gamma_0=0.57 \pm 0.13$ ($2\sigma$) for the Planck and WMAP values of $\Omega_{m0}$, respectively. 
 From our reconstruction, the growth index is more effectively constrained at $z = 0.09$, i.e., $\gamma=0.57 \pm 0.11$ ($2\sigma$), assuming the interval of $\Omega_{m0}$ given by the Planck collaboration. For the WMAP-9 estimate of the matter density parameter, we found a very similar value at $z = 0.05$. Note also that the current precision of the $H(z)$ measurements is not enough to place significant constraints on the $\gamma_0'$ value, which could provide a test of the $\Lambda$CDM model~\cite{GannoujiPolarski}. The final reconstruction of the growth index is presented in Fig. \ref{gammaz}.

\section{Conclusions}
\label{C}

In this work we have performed a non-parametric reconstruction of the cosmic expansion with cosmic chronometer and high-$z$ quasar data using the method of Gaussian Process. As discussed in Ref. \cite{Licia} the cosmic chronometer data until $z \sim 1.2$ are independent of cosmological and stellar population models. 
We have followed Ref.~\cite{Sahni2009} and calculated the most representative perturbative quantities in the GR frame with non-clustering DE, assuming spatial homogeneity and isotropy. For the values of $\Omega_{m0}$ given by the current CMB experiments, we have found a good agreement between current growth rate measurements  and the growth rate reconstructed  using the $H(z)$ data displayed in Table I (see {{ Fig. \ref{fz}}}). In other words, this amounts to saying that no evidence for a deviation from the standard cosmological description has been found in our analysis. On the other hand, a direct comparison of the reconstructed functions ($g(z)$, $f(z)$ and $\gamma(z)$) assuming different values of the matter density parameter clearly show the significant influence of this quantity in the calculations of the matter perturbations. 

We have also derived the value of the {{growth index at the present epoch, i.e., $\gamma_0=0.56 \pm 0.12$}} ($2\sigma$), whose evolution is almost constant until $z=1$. Such a result is compatible with the $\Lambda$CDM expected value $\gamma=0.545$  and with its first derivative $\gamma'_0 \simeq-0.015$ \cite{GannoujiPolarski}. Finally, we have shown that the reconstruction from the subsample of $H(z)$ data used in our analysis constrains the growth index to the interval  $0.51<\gamma(z)<0.62$ (1$\sigma$) at $z = 0.09$. Using a different approach and assuming a constant growth index, Ref.~\cite{Nesseris2008} found $0.505 < \gamma < 0.869$ (1$\sigma$).

\begin{acknowledgments}
JEG and JSA thank CNPq, CAPES and FAPERJ (Brazilian agencies) for the grants under which this work was carried out. JCC is supported by the DTI-PCI
program of the Brazilian Ministry of Science, Technology and Innovation (MCTI).
\end{acknowledgments}

\end{document}